# Radiation-tolerant high-entropy alloys via interstitial-solute-induced chemical heterogeneities

Zhengxiong Su[a,1], Jun Ding[b,1], Miao Song[c], Li Jiang[d], Tan Shi[a], Zhiming Li[e*], Sheng Wang[a*], Fei Gao[c], Di Yun[a], Chenyang Lu[a*], En Ma[b]

[a] Department of Nuclear Science and Technology, Xi'an Jiaotong University, Xi'an, 710049, China
[b] Center for Alloy Innovation and Design (CAID), State Key Laboratory for Mechanical Behavior of Materials, Xi'an Jiaotong University, Xi'an 710049, China
[c] Department of Nuclear Engineering and Radiological Sciences, University of Michigan, Ann Arbor, MI 48109, United States
[d] School of Materials Science and Engineering, Dalian University of Technology, Dalian, 116024, China
[e] School of Materials Science and Engineering, Central South University, Changsha, 410083, China

[1] These authors contributed equally to this work.

**High-entropy alloys (HEAs) [1-3] composed of multiple principal elements have been shown to offer improved radiation resistance[4-7] over their elemental or dilute-solution counterparts. Using NiCoFeCrMn HEA as a model, here we introduce carbon and nitrogen interstitial alloying elements[8,9] to impart chemical heterogeneities in the form of the local chemical order (LCO) and associated compositional variations. Density functional theory simulations predict chemical short-range order (CSRO) (nearest neighbors and the next couple of atomic shells) surrounding C and N, due to the chemical affinity of C with (Co, Fe) and N with (Cr, Mn). Atomic-resolution chemical mapping of the elemental distribution confirms marked compositional variations well beyond statistical fluctuations. Ni$^+$ irradiation experiments at elevated temperatures demonstrate a remarkable reduction in void swelling by at least one order of magnitude compared to the base HEA without C and N alloying. The underlying mechanism is that the interstitial-solute-induced chemical heterogeneities roughen the lattice as well as the energy landscape, impeding the movements of, and constraining the path lanes for, the normally fast-moving self-interstitials and their clusters. The irradiation-produced interstitials and vacancies therefore recombine more readily, delaying**

**void formation. Our findings thus open a promising avenue towards highly radiation-tolerant alloys.**

Our new strategy is to artificially introduce chemical heterogeneities, closely spaced with a length/size scale that is specifically suited for influencing point defect evolution, to enhance the irradiation tolerance of HEAs. To this end, we exploit interstitial alloying elements (i.e., C and N), which have varying chemical affinities with the principal elements (i.e., Ni, Co, Fe, Cr and Mn). The spatially dispersed C and N solutes accentuate LCO and associated sub-nanometer-scale variation in chemical composition, but not yet result in second-phase precipitates. The HEA we designed has a nominal composition of $Ni_{19.8}Co_{19.8}Fe_{19.8}Cr_{19.8}Mn_{19.8}C_{0.5}N_{0.5}$ (hereafter referred to as NiCoFeCrMn-CN). The actual chemical compositions of the bulk alloy measured using wet-chemical analysis are listed in **Supplementary Table 1**. **Supplementary Fig. 1a** presents the X-ray diffraction (XRD) pattern, revealing that the NiCoFeCrMn-CN is still a single-phase face-centered cubic (fcc) HEA, just as the base Cantor alloy NiCoFeCrMn (see **Supplementary Fig. 1b** for its grain structure examined using electron back-scattering diffraction (EBSD) mapping).

**Fig. 1** shows the elemental distribution in the NiCoFeCrMn-CN HEA after homogenization heat treatment at 1200 °C for 2 h (see **Methods**). Atom probe tomography (APT) in **Fig. 1a** indicates that no apparent carbides and nitrides can be observed in the 3D reconstructions. Due to the limited spatial resolution [9-11] of the APT technique, the distributions of all elements seem uniform. Therefore, to probe atomic-level chemical heterogeneities, we resorted to atomic-resolution energy-dispersive X-ray spectroscopy (EDS) mapping in a high-resolution aberration-corrected scanning transmission electron microscope (STEM), in combination with the DFT-based Monte Carlo simulations (**see Methods**). The TEM samples were prepared using focused ion beam (FIB) method. A flash electrochemical polishing technique was applied on FIBed foils to obtain ultra-thin and clean samples for atomic-resolution EDS mapping (see **Methods**).

The development of LCO is expected around C and N, as they have high affinity with some, but not all, of the principal elements in the base HEA. To see the preference for bonds that gives rise to the LCO (especially the CSRO in the first and second nearest neighboring atomic shells), we employed Monte Carlo simulations modeling based on a DFT-constructed lattice. The results indicate obvious chemical ordering, showing preferentially increased (Co, Fe) around C atoms, and (Cr, Mn) around N atoms, as shown in **Fig. 1b**. In addition, the atomic distribution of the first and second nearest neighbors between C and N atoms as a unit, as shown in **Supplementary Fig. 2**, respectively, and both exhibit the trend of Ni-poor and Co-rich. This is also expected to incur a change in the atomic fractions in the immediate vincinity of each interstitial solute, away from the sample-average compositon. In other words, the chemical composition becomes heterogeneous, varying from one location to another at a spacing corresponding to the distance between the individual C and N solutes.

**Fig. 1c** shows the atomic-resolution high-angle annular dark-field (HAADF) images and the corresponding EDS mapping of the NiCoFeCrMn-CN alloy with the [110] zone axis parallel to the electron beam. These maps reveal that each of the principal elements in this HEA distributes in an inhomogeneous fashion, on atomic level. A close examination, as circled in these EDS maps, suggests a number of brighter/dimer regions that stand out from the surroundings, with an approximate size of 0.5 nm in radius and a spacing on the order of 1-3 nm. Further enlarged images of such local regions (inside the bold white circles) are displayed in the top panelin **Fig. 1c**.In comparison, the base HEA appears to be more uniform, as seen from the atomic EDS mapping for NiCoFeCrMn shown in **Fig.1f** (from **Ref**. Qian Yu[12]. Reprinted from Springer).

**Fig. 1d** presents the line profiles in a (111) plane projected along the [110] zone axis taken from the EDS maps. We observe marked undulation (from 2% to 60%) in the atomic concentration of Ni, Co, Fe, Cr or Mn, with peaks and valleys appearing periodically at a spacing of the order of 1 nm. The period of this composition variation can also be assessed from the spatial autocorrelation length of each species, by using the *S* parameter analysis discussed in **Ref**. Qian Yu (see **Supplementary Fig. 3**). The resultant "wavelength" turns out to be consistent with the separation distance between individual interstitial solutes, which is expected to be about 1-3 nm for the 0.5 at. % C and 0.5 at. % N added HEA. The compositional variation in **Fig. 1d** is much more pronounced than that observed in the base HEA without C and N (the column-to-column statistical fluctuation in the NiCoFeCrMn Cantor alloy was reported to be ± 5%[12], as shown in **Fig.1g** (from **Ref**. Qian Yu[12]. Reprinted from Springer). As such, both the EDS mapping and the simulated CSRO point to appreciable chemical heterogeneity on sub-nanometer scale in NiCoFeCrMn-CN, markedly enhanced over that in the base Cantor alloy without the C- and N- additions.

Such wide-spread chemical heterogeneities are expected to cause variable local atomic bonding enviroments and therefore affect the evolution of point defects, which in turn alter the evolution of dislocation loops and void swelling. To observe the effect on irradiation response, we performed 3 MeV $Ni^{2+}$ irradiation at several elevated temperatures (420℃, 480℃, 540℃ and the peak dose about 55 dpa) . **Fig. 1e and Fig. 1h** present TEM images (the BF images were taken under focus to reveal the presence/distribution of voids) to compare NiCoFeCrMn-CN and NiCoFeCrMn after irradiation at 420℃. A striking difference is seen here:with C and N interstitial-solute-induced chemical heterogeneities, the radiation damages seen in NiCoFeCrMn, i.e., the obvious white voids, no longer appear in the NiCoFeCrMn-CN.

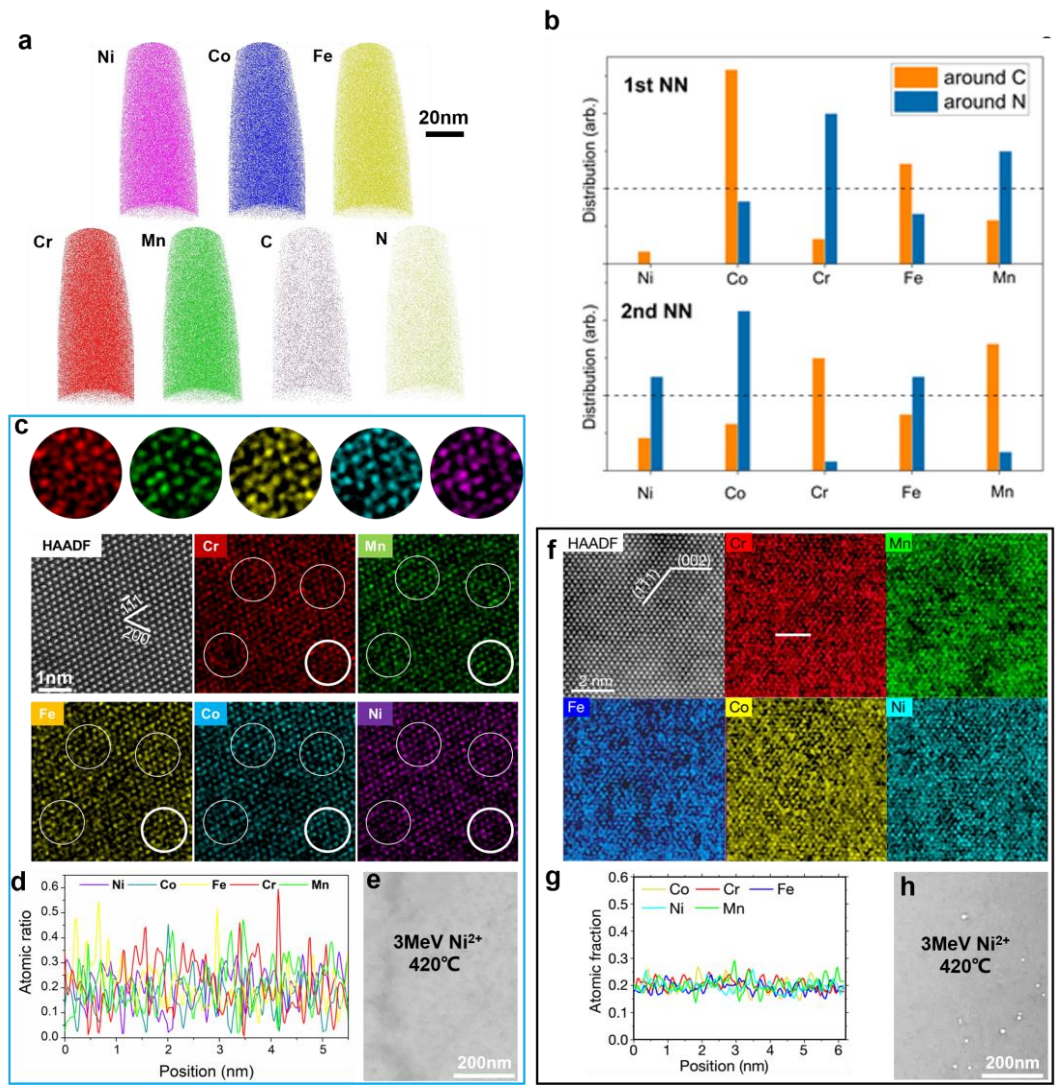

**Fig. 1 High-resolution mapping of elemental distributions and analysis of local chemical order in the NiCoFeCrMn-CN HEA. a**, Three-dimensional reconstructions

of a typical atom probe tomography (APT) tip. **b,** DFT-based MC simulation shows the distribution of principal elements around the carbon and nitrogen interstitial atoms in the first and second nearest neighbor (NN) shells; the dashed lines indicate the level corresponding to an ideally random solid solution. **c,** STEM-HAADF image of atomic structure, taken with the [110] zone axis parallel to the electron beam, and corresponding atomic-resolution EDS mapping for individual principal elements of Ni, Co, Fe, Cr, and Mn. The white circles highlight some brighter/dimmer sub-nanometer regions, with an enlarged view of a local region (inside the bold white circle) displayed in the top panel of **c**. **d,** Line profiles of atomic ratio of individual elements taken from the respective EDS mapping in **c**; each line profile represents the distribution of an element in a (111) plane projected along the [110] zone axis. **e,** Under-focused cross-sectional TEM images in NiCoFeCrMn-CN irradiated with 3 MeV Ni$^{2+}$ ions of peak dose about 50 dpa at 420 ℃. For a direct comparison with the base alloy without C and N, **f, g, h,** display the corresponding information (atomic-scale mapping and compositional fluctuations, from **Ref**. Qian Yu[12]. Reprinted from Springer), as well as irradiation induced voids) for NiCoFeCrMn.

    Next, we show in **Fig. 2a** cross-sectional BF TEM images of the samples irradiated at the three temperatures. Again, void formation is absent in the samples irradiated at 420 ℃ and 480 ℃. Apparently, the void is still in the incubation stage at these two relatively low temperatures, where the irradiation-induced free vacancies form stacking fault tetrahedra via a vacancy cluster diffusion and agglomeration mechanism[13] **(see Supplementary Fig. 4)**. In the sample irradiated at a much higher temperature (540 ℃), some voids show up, but are sparse and nonuniformly distributed, almost all of which are in the SRIM predicted ion-damaged range (<1600 nm). The degrees of void swelling of some typical fcc conventional and medium/high entropy alloys under similar irradiation conditions are compared in **Fig. 2b**[14-19]. Obviously, NiCoFeCrMn-CN is far superior to other conventional alloys, and also shows significantly better swelling resistance than the base HEA-NiCoFeCrMn. Although the irradiation was not performed at exactly the same temperature, this does not prevent us from concluding that the overall swelling (0.015%) in NiCoFeCrMn-CN is at least an order of magnitude lower than that (about 0.2%) of NiCoFeCrMn[14]. The interpolation of experimental data between 500 ℃ and 580 ℃ indicates that the onset of void formation has been significantly delayed to much higher irradiation temperatures. Moreover, both void growth and swelling are dramatically suppressed. Specifically,

thermally activated vacancy migration increases with increasing irradiation temperature, such that some vacancies would relocate beyond the predicted damage region, forming small voids in NiCoFeCrMn. In contrast, in NiCoFeCrMn-CN, voids are not observed beyond the ion range at 540℃, reflecting the restricted mobility of vacancy clusters to escape from the damage region[5,20]. The sparse voids found in the shallow region (< 600 nm) are small and isolated, indicating that the mobility of vacancy cluster in NiCoFeCrMn-CN is not as pronounced as that of NiCoFeCrMn. Vacancies appear to have a hard time escaping to deeper regions to aggregate into voids, which are therefore small, few and far between. The observation that the diffusion distance of vacancies is shorter in NiCoFeCrMn-CN than that in NiCoFeCrMn can be explained by the reduced vacancy migration rate in the presence of C and N, rendering the accumulation of vacancies difficult.

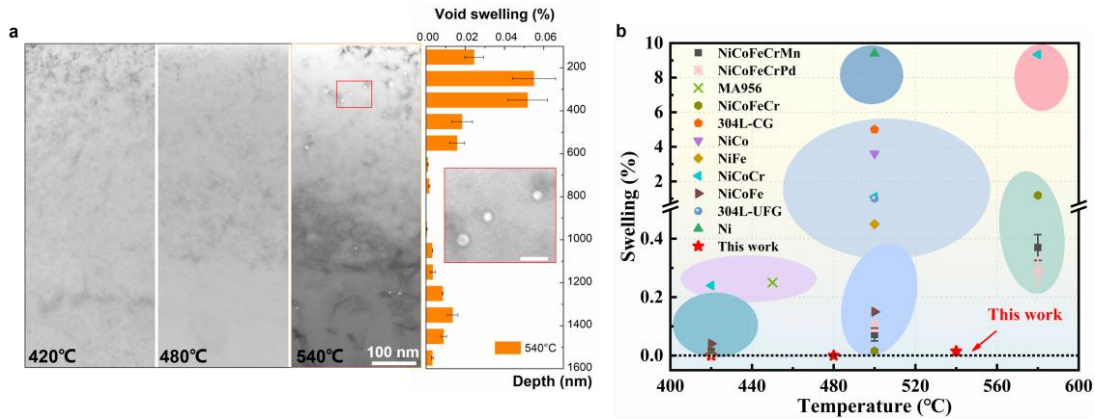

**Fig.2 Distribution of irradiation-induced voids. a,** Under-focused cross-sectional TEM images in NiCoFeCrMn-CN irradiated with 3 MeV $Ni^{2+}$ ions at a fluence of $5\times 10^{16}$ $cm^{-2}$ at 420℃, 480℃, and 540℃, respectively. The diagram on the right shows the void swelling distribution at various depths at 540℃. The scale bar in the zoomed images is 20 nm. **b,** A summary comparing the total void swelling versus irradiation temperature in various alloys[14-19]. All samples were probed with a peak dose of about 50 dpa. Some traditional alloys are far inferior to HEAs in their ability to tolerate swelling.

**Fig. 3a** displays the cross-sectional images, showing dislocation loop distributions in NiCoFeMnCr-CN after irradiation at 420℃, 480℃ and 540℃, respectively. The SRIM-calculated damage and injected ion concentration profiles are also shown in the left panel of **Fig. 3a**. All micrographs were taken under two-beam kinematical bright-field (BF) conditions using a diffraction vector **g** =200. Note that at 540℃, in contrast to the other two irradiation temperatures, some dislocation loops can be found well beyond the predicted damage region. This indicates that the irradiation-induced interstitial clusters have the ability to diffuse deep into the specimen only when the temperature is very high. In order to avoid the effects associated with the surface sinks, we examined in detail a region 300-500 nm in depth from the surface, irradiated to a dose of about 30 dpa, as shown in **Fig. 3b**. It can be observed that as the irradiation temperature increased from 420℃ to 540℃, the loop size increased moderately and its density rapidly decreased. **Fig. 3c** shows a STEM-BF image of entangled dislocation loops in the NiCoFeCrMn-CN irradiated at 540℃, and the faulted 1/3 <111> loops and perfect 1/2<110> loops are marked using yellow and red arrows, respectively. The half atomic plane is marked with a "T" sign, as shown in **Fig. 3d**, suggesting that this is an interstitial type 1/3<111> faulted loop, lying parallel to the incident electron beam. **Fig. 3e** shows that with increasing irradiation temperature from 420℃ to 540℃, the average density of dislocation loops decreases from $12.5×10^{21}$ m$^{-3}$ to $2.5×10^{21}$ m$^{-3}$, while their average size increases from 30.2 nm to 56.4 nm. It is worth noting that compared to the cases at 480℃ and 540℃, dislocation loops in the region around the implanted Ni concentration peak at 420℃ exhibit an interesting feature: the loop density is higher and the loop size is smaller, relative to those in NiCoFeCrMn. More detailed information on the evolution of defects with temperature can be found in **Supplementary Fig. 5**. This can be attributed to the marked reduction in the self-interstitial cluster mobility in NiCoFeCrMn-CN as compared to NiCoFeCrMn[14] (to be discussed further later).

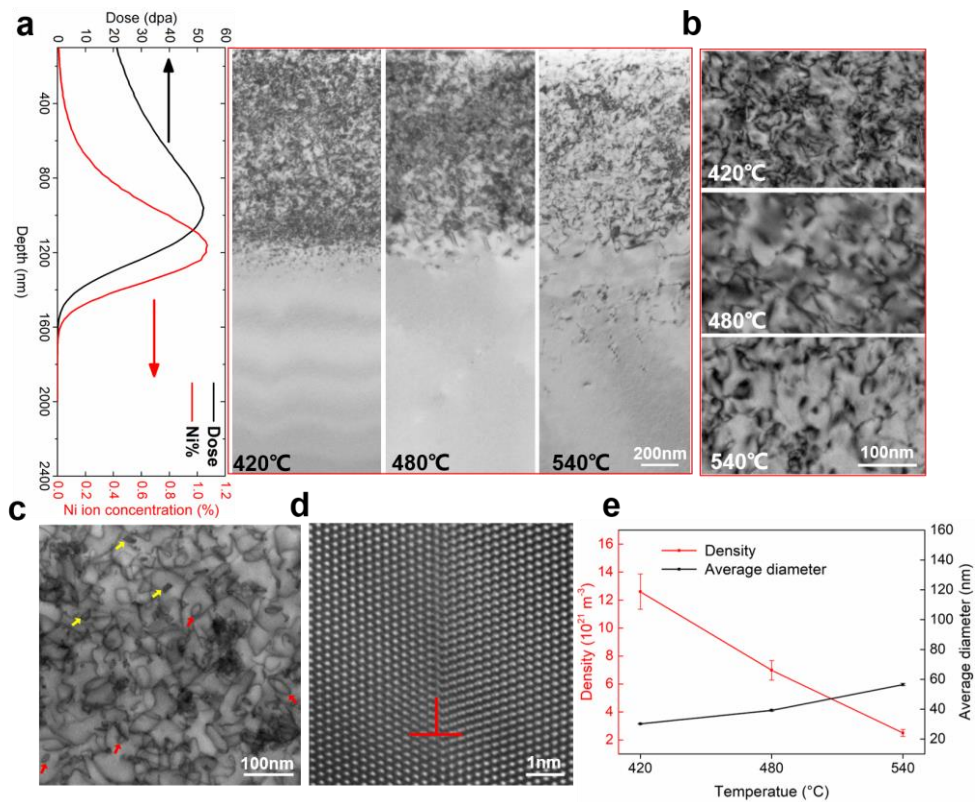

**Fig. 3 Distribution of irradiation-induced dislocation loops. a,** Cross-section TEM images showing the dislocation loop-depth distributions in NiCoFeCrMn-CN irradiated by 3 MeV $Ni^{2+}$ ions at a fluence of $5\times10^{16}$ $cm^{-2}$ at 420 ℃, 480 ℃ and 540 ℃, respectively. The depth profiles of damage and implanted Ni concentration calculated using SRIM-2013 software are shown on the left. **b,** Characteristics of dislocation loops irradiated to 30±5 dpa at 420 ℃, 480 ℃ and 540 ℃, respectively. **c,** STEM-BF image, taken from the [110] fcc zone axis, of two types of dislocation loops. Red arrows mark perfect loops and yellow arrows faulted loops. **d,** High-resolution HAADF image indicating the interstitial type of the dislocation loop. **e,** Average size and density of the dislocation loops.

We now discuss the mechanisms responsible for the unusual defect evolution, particularly the impressive void swelling resistance exhibited by the NiCoFeMnCr-CN alloy. Chemical heterogeneities have been reported to appreciably affect the dislocation behavior in HEAs, by increasing the dislocation core energy, pinning dislocations, and promoting their cross-slip and multiplication, which in turn improve the work hardening capability and the ductility of HEAs[12,21-24]. We project that chemical heterogeneities would also affect the evolution of dislocation loops and thus the radiation response in HEAs. As mentioned above, chemical heterogeneities in this work involve LCO and associated compositional variation: they are expected to go hand in hand as chemical short-range order inevitably instigate compositional deviations at the atomic level away from the nominal sample average.

The roles played by these chemical heterogeneities are two-fold. First, in the early stage of the irradiation, the chemical heterogeneities can effectively affect electron-phonon coupling to influence the energy dissipation behavior. The chemical heterogeneities would reduce the amplitude of $d$ electron hopping in the spin majority channel between the nearest atomic pair [25,26], thus decreasing the electron mean free path and changing the energy and mass transport properties. Furthermore, the spatial distribution of chemical heterogeneities at the nanoscale causes force variations all the time and all over the place, which give rise to significant phonon scattering and broadening[27]. Both of these phonon and electron effects contribute to the substantial reduction of thermal conductivity in the system, slowing down the dissipation of the deposited heat and prolonging the thermal spike. These effects enhance the defect recombination and disfavor void formation[28,29].

Second, the chemical heterogeneities reduce the diffusion rates of point defects. On the one hand, the chemical heterogeneities can induce local lattice distortion with associated atomic-level stress field fluctuations. Zhang *et al.* have shown that, according to their EXAFS analysis, the local lattice distortion in NiCoCr derives mainly from the LCO-induced preferable bonding between certain atomic pairs rather than from atomic size mismatch[30,31]. According to our simulation, the self-interstitial diffusion efficiency of NiCoFeCrMn-CN compared to that of NiCoFeCrMn is reduced by 36%. On the other hand, the LCO in NiCoFeCrMn-CN can further reduce the interstitial diffusivity by 34% (**Fig. 4a**). As a result of the sluggish diffusion of self-interstitial atoms, all processes in which interstitial atoms participate should be delayed, including those during dislocation loop growth. According to the defect reaction model[32,33], we know that the probability of annihilation between irradiation-induced vacancies and self-interstitials is maximum when their diffusion rates are similar, as that affords the best opportunities for them to run into each other and recombine. With the addition of C and N interstitials, the diffusion rate of self-interstitials is slowed down towards that of vacancies, as shown by the calculation results in **Fig. 4b**. This trend is more pronounced ($D_{vac}/D_{inter} \approx 0.1$) when LCO is elevated via annealing/ageing. As a consequence, the recombination probability of self-interstitial atoms and vacancies is effectively enhanced, which leads to a reduction in residual defect concentration and void swelling of the irradiated material.

Next, we explain why the migration path of self-interstitial clusters gets strongly altered due to the local chemistry variations[34]. The interaction between LCO and self-interstitial clusters is schematically shown in **Fig. 4c**. With increasing LCO, the atomic and energy landscapes become rougher, making the migration of point defects, and the growth of dislocation loops and voids, more difficult. This is because chemical heterogeneities will minimize the local free energy. They thus act like local trapping sites of the moving species, by raising their effective migration energy. Recent studies have revealed that LCO at the nanoscale intensifies the ruggedness of the energy landscape and increases the activation barriers for dislocation motion[22]. Similarly, the ordered chemical heterogeneities are expected to escalate the complexity of the energy landscape encountered by the migrating point defects. As a result of the roughened landscape, low-energy diffusion paths are interrupted, limiting the mobility of self-interstitials to delay their accumulation and long-distance travel[35]. The self-interstitial clusters thus have more opportunities to meet and take away irradiation-induced vacancies, delaying the coalescence of the latter and hence void nucleation. From the energy cost perspective, the diffusion of self-interstitial atoms would break the preferred LCO, rendering the local environment akin to "antiphase boundary". This incurs an energy penalty on the migration action, as demonstrated in **Fig. 4a**. In other words, the diffusion of self-interstitial clusters amidst chemical heterogeneities is like navigating across a rugged terrain, with frequent changes of direction required to seek the minimum migration energy path, rather than running smoothly as in a conventional solid solution alloy.

In this context, the nanoscale chemical heterogeneity is akin to the coherent "precipitates" in the alloy, but a notable difference is that the alloy here remains a single-phase solid solution [9,36]. Recent studies have found that the nanoscale chemical fluctuation could tweak the dislocation slip mode in HEAs from planner slip to wavy slip. The local chemically ordered (O, Ti, Zr)-structure can act as a tripping and re-routing obstacle for the moving dislocations, leading to frequent cross-slip and dislocation interactions, thereby improving the strain hardening and ductility of $(TiZrHfNb)_{98}O_2$ alloys[9]. Similar trapping effects have also be seen in NiCoFeCrPd, in which the introduction of Pd brings about a significant nanoscale composition fluctuation[12], slowing down the dislocations. Now that we are dealing with even smaller moving defects, i.e., the radiation-induced self-interstitial clusters, the effective "obstacle/trap" can be even smaller in size, to the point that even chemical short-range order and atomic-scale composition variations suffice to influence the migration mode of the point defects. Diffusing point defects are expected to be diverted with increasing chemical heterogeneity. As the migration paths are no longer equivalent in all directions due to the variability in the local chemical composition/order, straight-ahead hopping of self-interstitial atoms or vacancies tends to become "dancing around in circles". In previous work adopting the multi-principal element recipe[37], we have observed that the migration paths of defects change from random long-range 1D-diffusion to local short-range 3D-diffusion. The same is true for dislocation loops[38]; an in-situ experiment confirmed that in a highly concentrated solution, the jump frequency and the glide distance of the dislocation loops were reduced, because destroying chemically inhomogeneous structure requires additional energy and generates a local antiphase boundary. By the same token, in NiCoFeCrMn-CN the migration of the radiation-induced defects would be hampered, such that their paths become tortuous and their transport confined in local regions, as depicted in **Fig. 4c.**

Finally, we mention in passing that the changed mode in defect evolution, in the presence of heightened chemical heterogeneity, is also reflected by the hardening effect of the alloys due to irradiation-induced defects, as revealed in nanoindentation tests **(Supplementary Fig. 6)**. The nanoindentation hardness can be utilized as an effective indicator of the mechanical changes of the alloys before and after irradiation[39]. The effect of defect cluster on hardening is summarized in **Supplementary Table 2,** in which experimental and predicted hardness changes for the irradiated samples are shown to be consistent. Furthermore, the decrease in hardening with increasing irradiation temperature can be attributed to a sharp reduction in loop density and a slight increase in loop size, as the small interstitial clusters tend to migrate and accumulate together to form larger loops at higher irradiation temperatures. Overall, the more visible hardening effect of NiCoFeCrMn-CN compared to that of NiCoFeCrMn is due to delayed dislocation loop evolution. Besides, the dislocation loop density of NiCoFeCrMn-CN irradiated at 540℃ is similar to that of the NiCoFeCrMn irradiated at 420℃, indicating that the dislocation loop evolution is indeed slower after the addition of C and N.

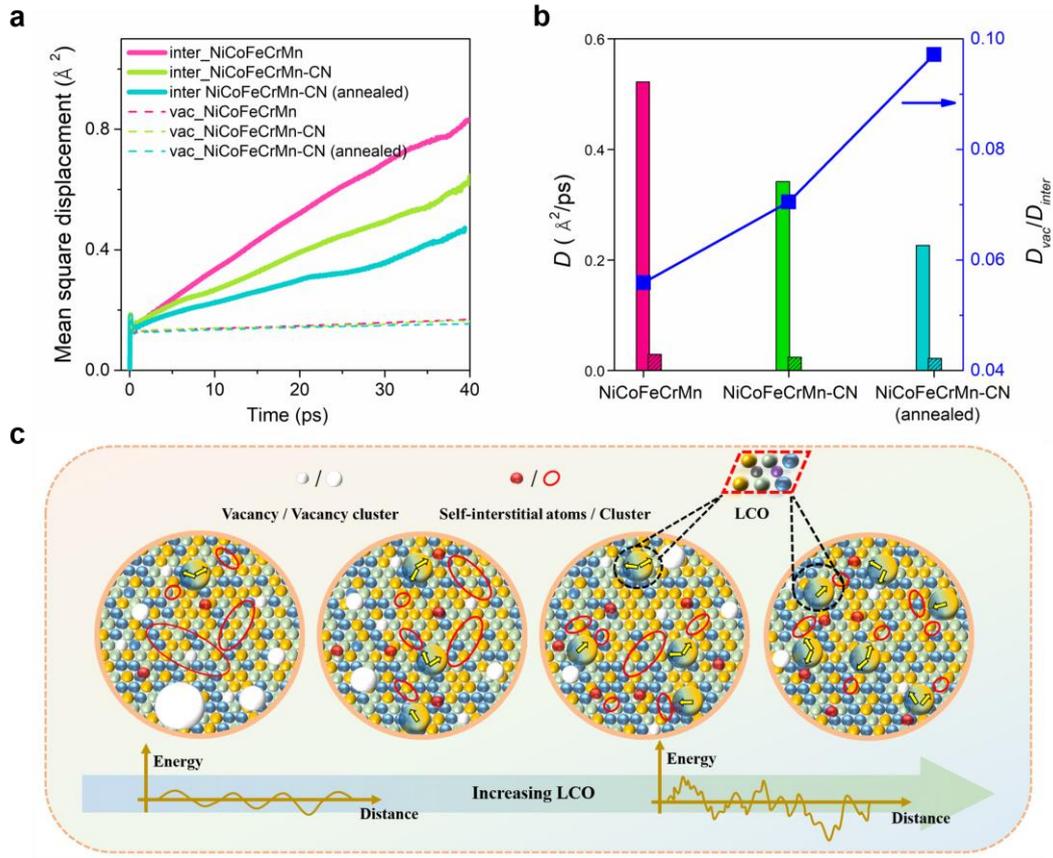

**Fig. 4 Local chemical order slows down diffusion. a**, The mean square displacement of vacancies and interstitial atoms in NiCoFeMnCr, NiCoFeMnCr-CN, and NiCoFeMnCr-CN (annealed), respectively. **b,** The diffusion coefficients of self-interstitial atoms ($D_{inter}$) and vacancies ($D_{vac}$), as well as their ratio. **c**, Schematic illustration of the interplay between LCO and radiation-induced defects in a FCC HEA. The arrows represent the defect is attracted by LCO and changes its migration direction. The bottom schematic illustrates the migration energy landscape confronting the defects.

In summary, the present study highlights the power of tailoring chemical heterogeneities as a new strategy to improve the radiation resistance of alloys. We have shown that the C and N interstitial solutes prompt local chemical order and composition variations, in addition to increasing lattice strain. The extra chemical heterogeneities raise the barriers against the diffusion of irradiation-induced interstitial atoms and vacancies, altering the rates and pathways of these defects in the direction of facilitated recombination. As a result, dislocation loops are harnessed and void formation is delayed. The radiation damage is therefore reduced. Our findings have important and broad implications: HEAs can be alloyed with other LCO/fluctuation-provoking interstitial or even substitutional elements. These purposely introduced chemical heterogeneities holds promise for rendering alloys with extended service time and expanded operating temperature range, both desirable for the cost and energy efficiency of nuclear reactors.

## Methods

### Alloy preparation

The HEA with interstitial carbon and nitrogen solutes, with nominal composition $Ni_{19.8}Co_{19.8}Fe_{19.8}Cr_{19.8}Mn_{19.8}C_{0.5}N_{0.5}$ (at. %), was prepared from pure metals, carbon, and $FeCrN_2$ (as the source of nitrogen) using a vacuum induction furnace [40]. The as-cast HEA was hot-rolled at 900 °C with a thickness reduction ratio of 50%, and then heat-treated at 1200 °C for 2 h in an Ar atmosphere. The HEA plate was water-quenched after the heat treatment. The chemical compositions of the resultant bulk alloy were measured using wet-chemical analysis, as listed in **Supplementary Table 1**. Samples machined from the HEA plate were ground using SiC paper (up to # 4000 grit), and then polished using a 0.05 μm alumina polishing solution to eliminate deformation layers resulted from mechanical grinding. "Mirror-like" surfaces were achieved to guarantee reliable results from nanoindentation and irradiation.

### Ion irradiation experiments

The samples were irradiated with 3 MeV $Ni^{2+}$ ions at a fluence of $5×10^{16}$ $cm^{-2}$ at 420 ℃, 480 ℃ and 540 ℃, respectively. The flux was controlled at $2.8×10^{12}$ $ions/cm^2/s$. A rastered beam was used to ensure the homogeneous ion irradiation. Predicted local dose and implanted Ni ion concentration were calculated using SRIM-2013 in Quick Kinchin-Pease Mode with a displacement threshold energy of 40 eV for all the constituent elements. The peak damage corresponded to a damage dose of ~50 displacement per atom (dpa) at the depth of 1,000 nm.

**Microstructural characterization**

The cross-sectional TEM samples were prepared using the focused-ion beam (FIB) lift-out technique using a FEI Scios 2 Dualbeam. A flash electrochemical technique was applied to remove FIB-induced damage caused by the $Ga^+$ ions bombardment to obtain ultra-thin and clean samples. The low-magnification characterization of voids and dislocation loops after irradiation was performed in a JEOL F200 TEM operated at 200 keV. A Cs-corrected S/TEM JEM-ARM300F2 operated at 300 keV was employed for high-resolution STEM-bright field (STEM-BF) and high angle annular dark field (HAADF) imaging, and corresponding atomic resolution EDS mapping. The statistics of irradiation-induced defect clusters were analyzed using an Image J software, starting from 100 nm in depth from the irradiation surface to avoid the strong surface effect. The size of the dislocation loops was counted by measuring their longest axis. The statistical error for voids follows the methods in Ref [41] and the void swelling value was calculated as the ratio of the volume of detectable voids to the analyzed sample volume [42]. Thousands of dislocation loops were counted in each irradiation condition to increase statistical accuracy. For the analysis of voids, four FIB samples with a total imaged area of 6 μm² in different regions were counted.

**Simulation methodology**

DFT-based Monte Carlo simulations were implemented for FCC NiCoFeCrMn-CN alloys. Four independent supercells of NiCoFeCrMn, containing 180 atoms each, were generated as special quasi-random structure (SQS)[43] as the initial starting points; then 2 C and 1 N atoms were randomly inserted into the octahedral sites as interstitials for each supercell. The temperature employed in the MC simulations was 600 K. Energy calculations were performed with the Vienna ab initio simulation package (VASP)[44-46], using spin-polarized density-functional theory, with a plane wave cut-off energy of 420 eV. Brillouin zone integrations were performed using Monkhorst–Pack meshes with a 2×2×2 grid[47]. Projector-augmented-wave (PAW) potentials[46] were employed with the Perdew-Burke-Ernzerhof (PBE) generalized-gradient approximation (GGA) for the exchange-correlation functional[48]. Similar to the methods used in **Ref. 24**, lattice MC simulations included swaps of atom types (only for metallic elements) with the acceptance probability based on the Metropolis–Hastings algorithm[49].

The *ab initio* molecular dynamics (AIMD) simulations were performed for three types of HEAs: solid-solution NiCoFeCrMn-CN, annealed NiCoFeCrMn-CN and NiCoFeCrMn. 4 independent supercells for each HEA were studied at 1,200 K with a fixed supercell volume, and the temperature was controlled by the Nose–Hoover thermostat. The energy cutoff of 300 eV was used with a single k-point (Γ). Spin polarization was not included, considering that the studied temperature in this experiment was much higher than the materials' Curie temperatures[50]. A single [100] interstitial dumbbell or vacancy was randomly introduced at one of the lattice sites (not close to C and N atoms) to initiate the simulations for interstitial[51] and vacancy diffusions, respectively. The defect structure (interstitial) was first equilibrated over 5 ps, and the diffusion data was then collected from the subsequent 40-ps simulations. By scrutinizing the atomic details, their diffusion coefficients (*D*), for vacancy and interstitial diffusion, were calculated as $D = \frac{<r^2(t)>}{6t}$, where $<r^2(t)>$ is the corresponding mean square displacement (MSD).

**Acknowledgments:** This work is supported by the National Key Research and Development Program of China under Grant No.2019YFA0209900 and 2017YFB0304403; the National Natural Science Foundation of China under Grant No. 12075179; We thank JEOL and I.Ohnishi for providing the JEOL-ARM300F2 used for atomic-scale EDS mapping.

# Supplementary materials

**Supplementary Table 1.** Chemical composition of NiCoFeCrMn-CN alloys

| Alloys | Ni | Co | Fe | Mn | Cr | C | N |
|---|---|---|---|---|---|---|---|
| Atomic % | 19. 64 | 20.12 | 17.92 | 19.67 | 21.21 | 0.60 | 0.83 |
| Weight % | 20.80 | 21.40 | 18.06 | 19.50 | 19.90 | 0.13 | 0.21 |

**Supplementary Fig. 1. a**, **b**, XRD pattern and electron back-scatter diffraction maps of the FCC solid-solution NiCoFeCrMn-CN HEA, respectively.

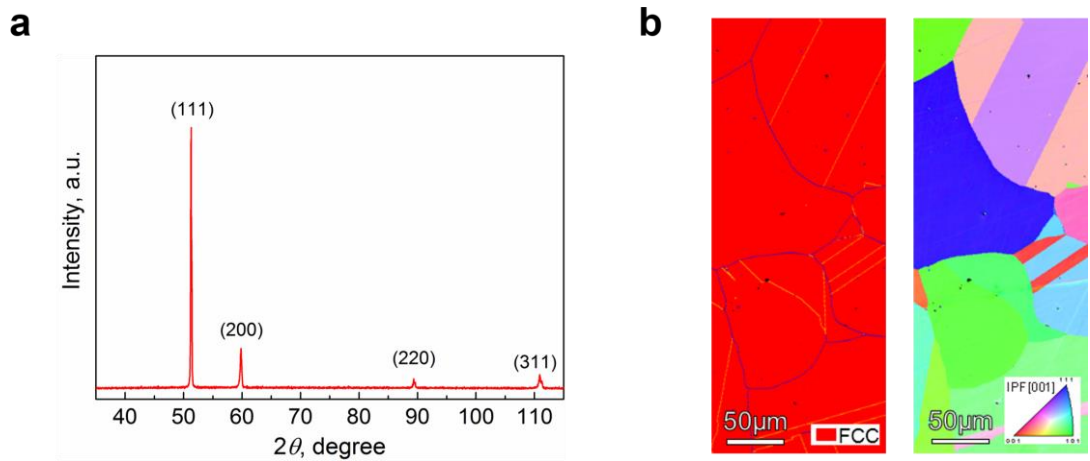

**Supplementary Fig. 2.** The atomic distribution of the first and second nearest neighbors between C and N atoms as a unit, respectively.

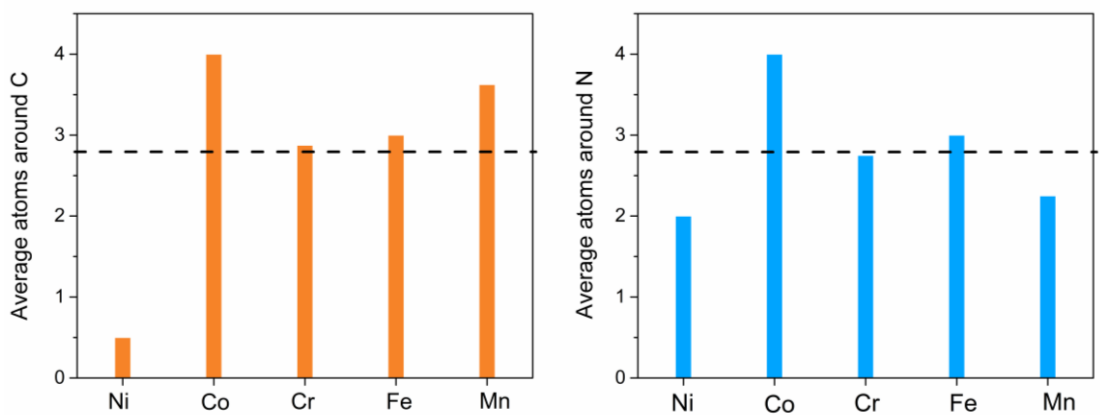

**Supplementary Fig. 3**. Plots of pair correlation function $S(r)$ of individual elements against concentration wavelength $r$ for (a) NiCoCrFeMn-CN and (b) NiCoCrFeMn; $S(r)$ is shifted by $C^2$, where $C$ denotes the average atomic fraction of the corresponding element. The data for NiCoCrFeMn are extracted from Q.Q. Ding et al. Nature. 574, 223 (2019).

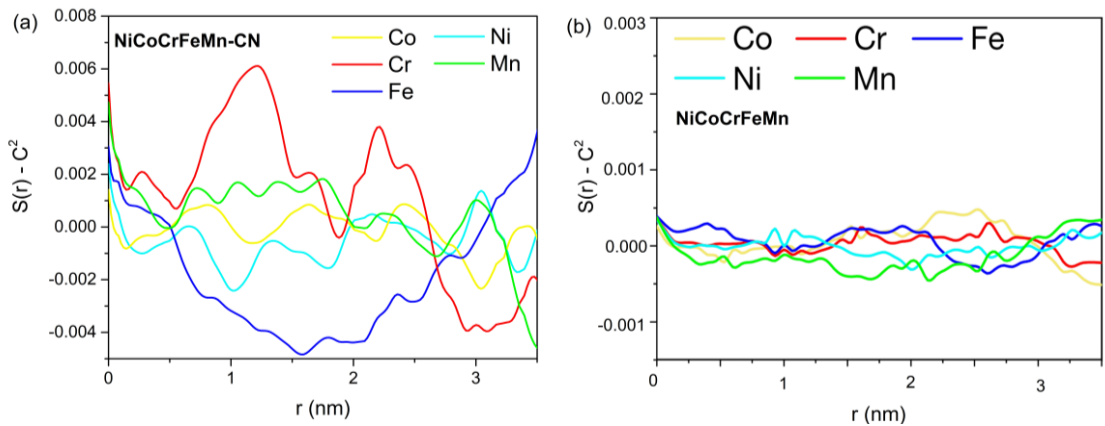

**Supplementary Fig. 4.** STEM images of vacancy-type stacking fault tetrahedra in the sample irradiated at 480 ℃. The SFT is visible as open triangles bordered by 111 planes, as highlighted by circles. The STEM-HAADF and STEM-BF images were taken with the electron beam direction parallel to [110].

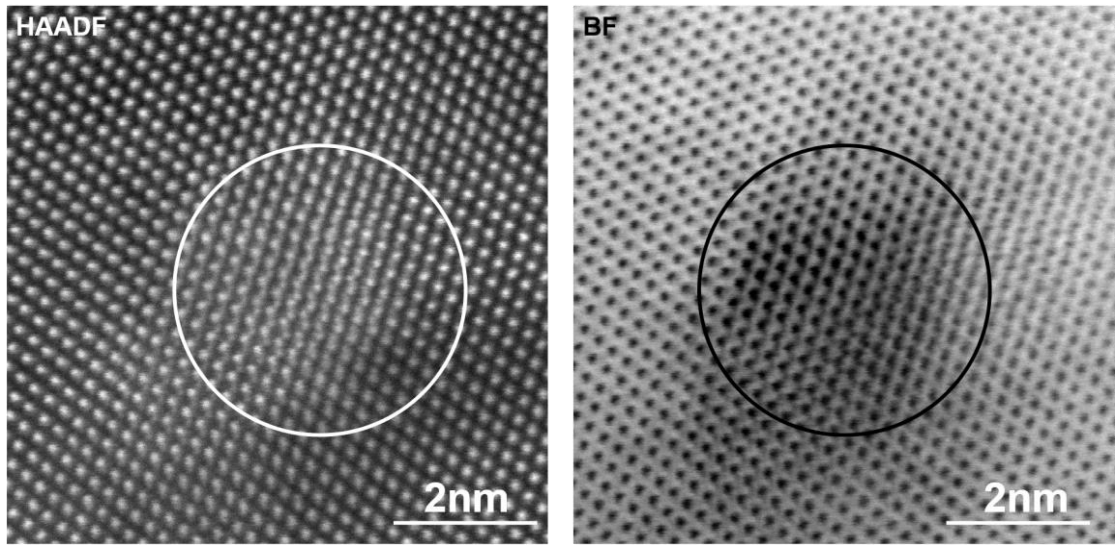

**Supplementary Fig. 5.** The statistical distribution of dislocation loops in irradiated NiCoFeCrMn-CN. Comparison of dislocation loop-depth distributions in NiCoFeCrMnCN irradiated with 3 MeV $Ni^{2+}$ ions to a fluence of $5\times10^{16}$ cm$^{-2}$ at 420℃, 480℃ and 540℃. **a,** Average size. **b,** Number density **c,** Size distribution

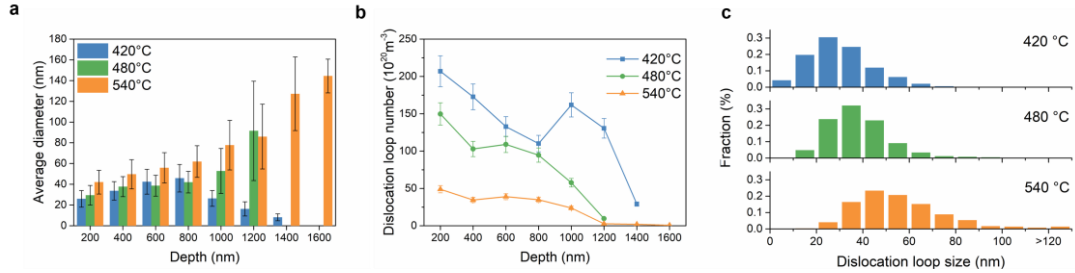

The irradiation temperature can significantly affect the formation and evolution of defects/defect clusters. However, the study on the dislocation loop behavior of HEAs in relation to irradiation temperature is rather limited. Based on **Supplementary Fig. 5,** with increasing irradiation temperature, the size of loops increased while the density of the loops decreased. Simultaneously the fraction of small loops gradually decreased and even disappeared at 540℃ because of the generation of large loops. The loop evolution process within increasing temperature can be attributed to the increasing interstitial mobility, dissociation of small interstitial clusters and reforming of the large interstitial clusters. Similar effects have been observed in previous FCC alloys[52,53].

The density and average size of dislocation loops are shown in **Supplementary Fig. 5a-b**. In 480℃ and 540℃ irradiated samples, with increasing the irradiation depth the size of dislocation loops increases while the density decreases. At the end of ion-damage depth, extended dislocation lines form due to the accumulation of dislocation loops. It should be noted that after irradiation at 420℃, the largest dislocation loop appeared at a depth of 800-1000 nm, with an average size of about 50 nm, and a sharp peak of density also arises in this region. In general, the size distribution of dislocation loop reflects the evolution of dislocation loops at three different irradiation temperatures. As shown in **Supplementary Fig. 5c**, with increasing temperature the fraction of smaller dislocation loop gradually decreases, while that of larger ones gradually increases. In addition, many larger loops (>70nm) appeared at 540℃ but not at the other two temperatures. The average loop size in the sample irradiated at 540℃ was 57 nm, which was about twice that at 420℃. In contrast, the density of the loops in the sample irradiated at 540℃ decreases sharply to $24.8\times10^{20}$m$^{-3}$, which is approximately seven times smaller than at 420 ℃.

In addition, very few voids are seen at the depths 600-1200 nm from the surface, forming a void-denuded region **(Fig. 2a)**. This observation can be attributed to the relatively high recombination rate of vacancy-interstitials in this region, resulting from the high density of self-interstitial atoms. Besides, the recombination process caused by the effect of implanted Ni is relatively minor due to the limited peak ion concentration and high density dislocation loops observed in the irradiated region[42].

**Supplementary Fig. 6.** Irradiation induced hardening effect. **a**, Load–unload curves of a pristine and an irradiated NiCoFeMnCr-CN. **b**, Hardness and irradiation induced hardening.

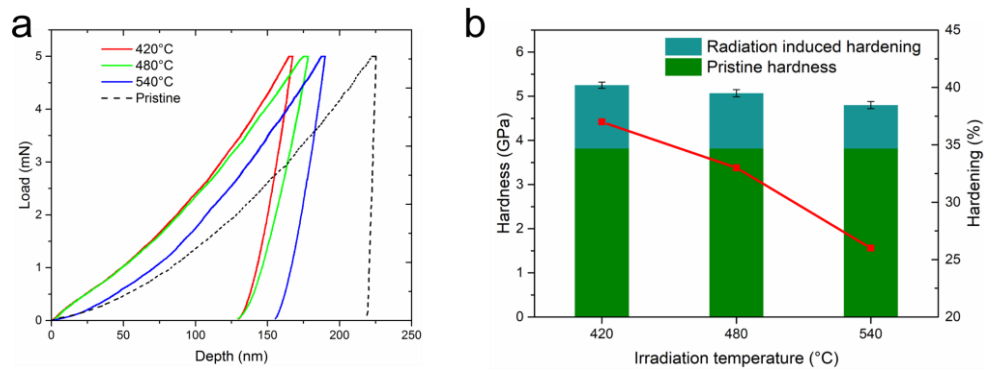

A Hysitron TI 950 Triboindenter was used for the tests using a Berkovich triangular pyramid indenter. A number of indentations were performed on both pristine and irradiated regions to achieve a better statistical analysis at a constant $P/P=0.05s^{-1}$. As shown in **Supplementary Fig. 6**, there are obvious differences in the load-unload curves between pristine and irradiated samples, indicating the change in hardness. The sample irradiated at 420 ℃ presented the most hardening effect, and the sample irradiated at 540 ℃ showed the least hardening effect.

**Supplementary Table 2.** The predicted increase in hardness compared with that measured in the nanoindentation experiment.

| Irradiation temperature (°C) | Defect type | Density ($10^{21}$/m³) | size (nm) | Calculated change in hardness (GPa) | Experimental change in hardness (GPa) |
|---|---|---|---|---|---|
| 420 | Dislocation loop | 12.5 | 30.2 | 1.47 | 1.43 |
| 480 | Dislocation loop | 6.9 | 39.2 | 1.24 | 1.25 |
| 540 | Dislocation loop | 2.5 | 56.5 | 1.06 | 0.98 |
|     | Void | 0.053 | 15.2 |  |  |

It is well known that the irradiation-induced defect clusters could act as obstacles to pin the moving dislocations, leading to hardening. The measured nanoindentation hardness is compared with the calculation based on the dispersed barrier hardening (DBH) model:

$$\Delta \sigma_y = M \alpha \mu b \sqrt{(ND)}$$

where $\Delta \sigma_y$ is the yield strength increase, M is the Taylor factor (3.06 for fcc metal), α is the obstacle strength factor (0.4 for dislocation loops, 1 for voids[39,54]), µ is the shear modulus (80 GPa for NiCoFeCrMn alloys), *b* is the module of the Burgers vector (0.257nm), N and D are the number densities and average size of defects, respectively. The yield strength and hardness is proportional to each other [55].

$$\Delta H = K \Delta \sigma_y$$

where the constant K is about 3 for irradiated austenitic stainless steels.

As a result of the changes in N and D, with increasing temperature the hardening effect gradually decreases due to the growth of dislocation loops. The hardening effect of NiCoFeMnCr-CN at 540℃ is therefore predicted to be similar to that of NiCoFeMnCr at 420℃, based on the dislocation loop population.